\documentclass[aps, pra, showpacs, amsmath, twocolumn]{revtex4}
\usepackage{graphicx}
\usepackage{dcolumn}
\usepackage{bm}
\usepackage{amsmath}

\begin{document}

\title{Quantum dynamics of repulsively bound atom pairs in the Bose-Hubbard model}
\author{Li Wang}
\affiliation{Beijing National Laboratory for Condensed Matter
Physics, Institute of Physics, Chinese Academy of Sciences,
Beijing 100080, China}
\author{Yajiang Hao}
\affiliation{Beijing National Laboratory for Condensed Matter
Physics, Institute of Physics, Chinese Academy of Sciences,
Beijing 100080, China}
\author{Shu Chen}
\email{schen@aphy.iphy.ac.cn}
\affiliation{Beijing National
Laboratory for Condensed Matter Physics, Institute of Physics,
Chinese Academy of Sciences, Beijing 100080, China}
\date{\today }

\begin{abstract}
We investigate the quantum dynamics of repulsively bound atom
pairs in an optical lattice described by the periodic Bose-Hubbard
model both analytically and numerically. In the strongly repulsive
limit, we analytically study the dynamical problem by the
perturbation method with the hopping terms treated as a
perturbation. For a finite-size system, we numerically solve the
dynamic problem in the whole regime of interaction by the exact
diagonalization method. Our results show that the initially
prepared atom pairs are dynamically stable and the dissociation of
atom pairs is greatly suppressed when the strength of the on-site
interaction is much greater than the tunneling amplitude, i.e.,
the strongly repulsive interaction induces a self-localization
phenomenon of the atom pairs.
\end{abstract}


%

\maketitle

\section{Introduction}

The ultracold bosonic atoms in optical lattices have opened a new
window to study many-particle quantum physics in a uniquely
controlled manner \cite {Bloch,Greiner,Stoeferle}. Various schemes
have been proposed to realize a wide range of many-body
Hamiltonians by loading ultracold atoms into a certain optical
lattice \cite{Jaksch2005,Jaksch,Duan}. The advances in the
manipulation of atom-atom interactions by Feshbach resonances
\cite{Inouye} allowed experimental study of many-body systems
accessible to the full regime of interactions, even to the very
strongly interacting Tonks-gas limit \cite{Kinoshita,Tonks}.
Recently a lot of exciting experiments in optical lattices have
been implemented, including the superfluid-Mott-insulator
transition \cite{Greiner,Stoeferle}, non-linear self-trapping in a
periodical optical lattice \cite{Anker} and repulsively bound atom
pairs in an optical lattice \cite{Winkler}. The basic physics of
the ultracold atomic systems in optical lattice is captured by the
Bose-Hubbard model (BHM) \cite{Jaksch,Jaksch2005}, which
incorporates the competition between the interaction and hopping
energy and has been successfully applied to interpret the
superfluid-Mott-insulator transition. As a fundamental model, the
BHM has been widely applied to study the quantum phase transitions
and dynamic problems in the optical lattices.

Very recently, Winkler \textit{et al.} \cite{Winkler} have studied
the dynamical evolution of the initially prepared bosonic atom
pairs in an optical lattice. Their experimental results indicate
that the atom pairs with strongly repulsive on-site interaction
$U$ exhibits a remarkably longer lifetime than the system with
weakly repulsive interaction \cite{Winkler}. At first glance, this
result is counter-intuition because one may expect the repulsive
interaction to separate the particles, instead to bind them
together. The experimental result has stimulated theoretical
investigations on the dynamics of repulsively bound pairs
\cite{Denschlag,Petrosyan}. In Ref.\cite{Winkler}, the theoretical
understanding of the stable pair relies on the analytical solution
of a two-particle problem by solving two particle
Lippmann-Schwinger scattering equation on the lattice
corresponding to the Bose-Hubbard
Hamiltonian\cite{Winkler,Denschlag}. Obviously, this method is
only limited to a single repulsively bound pair and is not capable
to extend to deal with many-particle dynamic problem.

Motivated by the experimental progress \cite{Winkler}, in this
paper we study the quantum dynamics of the repulsively pair states
in the BHM both analytically and numerically. In the strongly
repulsive limit, we develop an analytical method to deal with the
dynamical problem based on the perturbation expansion of the
hopping terms, whereas we can numerically solve the dynamic
problem in the whole regime of interaction for a finite-size
system which could be diagonalized exactly by the exact
diagonalization method. The Bose-Hubbard Hamiltonian (BHH) reads
\cite{Jaksch,Fisher}
\begin{equation}
\hat{H}=-J\sum_{\left\langle i,j\right\rangle }\hat{b}_i^{\dagger
}\hat{b}_j+\frac 12U\sum_i\hat{n}_i(\hat{n}_i-1),  \label{BHH}
\end{equation}
where $\hat{b}_i^{\dagger }(\hat{b}_j)$ is the creation
(annihilation) operator of bosons on site $i$ $(j)$,
$\hat{n}_i=\hat{b}_i^{\dagger }\hat{b}_i$ counts the number of
bosons on the $i$th site, and $\left\langle i,j\right\rangle $
denotes summation over nearest neighbors. The parameter $ J $
denotes the hopping matrix element to neighboring sites, and
$U\propto a_s$ represents the on-site interaction due to s-wave
scattering. For an actual optical lattice, $J$ and $U$ are related
to the depth of the optical lattice $V_0$ which is determined by
the intensity of the laser. The lattice constant $a$ is half of
the wave length of the laser $\lambda /2$ \cite {Jaksch}. In this
article, we focus on the dynamical evolution of the repulsively
bound atom pairs in the periodic Bose-Hubbard model with $U>0$,
i.e. repulsive on-site interaction. In the following calculation,
we will ignore all possible dissipations in the system, such as
the loss of atoms by three-body collision.

The paper is organized as follows. In Sec. II, we first review a
general scheme to deal with dynamical evolution and present the
exact result for the two-site problem. In Sec. III, we develop a
perturbation method to study the dynamic evolution of the
initially prepared state of atom pairs which works in the large
$U$ limit. In Sec. IV, we study the dynamical problem for a
finite-size system by using the exact diagonalization method and
compare the analytical results with the exact numerical results.

\section{General scheme}

The Bose-Hubbard model has been investigated by a variety of
theoretical and numerical methods under different cases
\cite{Fisher,Stoof,Roth,Norman}. Most of the theoretical
investigations concern the ground state properties, whereas the
quantum dynamic problems are hardly dealt with and most of works
are limited to the double-well (two-site) problems
\cite{Vardi,Tonel}.

Given an initial state $\left| \psi _0\right\rangle ,$ the
evolution state at time $t$ can be formally represented as
\begin{equation}
\left| \psi (t)\right\rangle =e^{-i\widehat{H}t}\left| \psi _0\right\rangle .
\end{equation}
If we know all the eigenstates of the BHH which are the solutions
of schr\"{o}dinger equation $\widehat{H}\left| \phi
_n\right\rangle =E_n\left| \phi _n\right\rangle ,$ we can get
\begin{equation}
\left| \psi (t)\right\rangle =\sum_ne^{-iE_nt}\left| \phi
_n\right\rangle \left\langle \phi _n\right. \left| \psi
_0\right\rangle . \label{psit}
\end{equation}
Now it is straightforward to obtain the probability of finding the
given initial state at time $t$
\begin{eqnarray}
P_0(t) &=&\left| \left\langle \psi (0)\right. \left| \psi (t)\right\rangle
\right| ^2  \nonumber \\
&=&\sum_{n=1}^D\sum_{m=1}^D\left| a_n\right| ^2\left| a_m\right|
^2e^{-i(E_n-E_m)t}, \label{p0t}
\end{eqnarray}
where $a_n=\left\langle \phi _n\right. \left| \psi _0\right\rangle
$ and $ D=\left( N+L-1\right) !/[N!\left( L-1\right) !]$ is the
basis dimension of the $N$-particle system in a lattice with size
$L$. If the energy spectrum and its corresponding eigenstates
$\phi_n$ are known, then the dynamical problem is exactly solved
in principle. It is obvious that the dimension $D$ increases very
quickly as the particle number $N$ and lattice site $L$ increase.
Therefore it is not practical to investigate the dynamical problem
of a large system in this way, though one can solve it numerically
for a finite-size system by exact diagonalization method.
\begin{figure}[tbp]
\includegraphics[width=9cm]{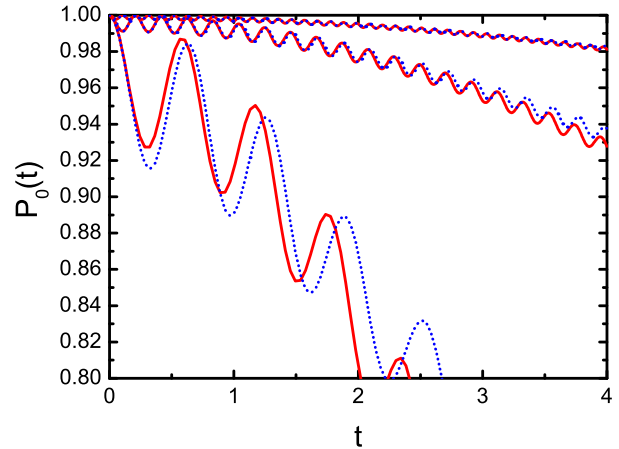}\newline
\caption{(Color online) Time evolution of $P_0(t)$ versus $t$ in
units of $J$ for the two-site-two-boson case. A comparison between
the exact and the approximate solution is made. The solid lines
denote the exact solutions given in Sec. II. The dotted lines
denote the approximate solutions by perturbation theory. From top
to bottom the lines correspond to $U/J=60$, $30$ and $10$,
respectively. } \label{Figure1}
\end{figure}

It is instructive to study the two-site system before presenting
the analytical results based on perturbation theory and the
numerical results for the systems with larger size. The dynamical
evolution of an atom pair in the toy two-site model can be
completely solved analytically. The results of this toy model will
shed light on studies of larger systems. According to the scheme
discussed above, it is quite easy to get the analytical expression
of $P_0(t)$ for a pair of atoms initially in a two-site lattice:
\begin{eqnarray}
P_0(t)&=&\left| a_1\right| ^4+\left| a_2\right| ^4+\left|
a_3\right| ^4 \nonumber \\
&&+ 2 \left| a_1\right|^2 \left| a_2\right|^2 \cos
\left(E_1-E_2\right ) t \nonumber\\
&&+2 \left| a_2\right|^2 \left| a_3\right|^2 \cos \left( E_2-E_3
\right) t \nonumber \\
&&+2 \left| a_3\right|^2 \left| a_1\right|^2 \cos \left( E_3-E_1
\right ) t \label{2exact},
\end{eqnarray}
where $a_1=1/\sqrt{2}$, $a_2=1/\sqrt{2+\alpha^2}$ and
$a_3=1/\sqrt{2+\beta^2}$ with the parameters $\alpha=-4 \sqrt{2}
J/\left( U-\sqrt{U^2+16 J^2} \right)$ and $\beta=-4 \sqrt{2}
J/\left( U+\sqrt{U^2+16 J^2}\right)$. The eigenenergies are given
by $E_1=U$, $E_2=\left(U-\sqrt{U^2+16 J^2}\right)/2$, and
$E_3=\left(U+\sqrt{U^2+16 J^2} \right)/2$. It is quite easy to
show the time evolution of the initial atom pair by the formula
(\ref{2exact}). In Fig. 1, we plot the time evolution of $P_0(t)$
for $U/J=60$, $30$ and  $10$. The results show that the larger the
repulsive $U$ is, the probability of finding the initial atom pair
is larger. When $U/J \rightarrow \infty$, we observe that $a_2
\rightarrow 0$, $a_3 \rightarrow 1/\sqrt{2}$, $E_1 - E_2 \approx
E_3 - E_2\approx U$ and $E_3 - E_1 \approx 4 J^2/U$. In this
limit, the terms including of $a_2$ in Eq. (\ref{2exact}), which
is proportional to $ \frac{J}{U} \cos (Ut) $, merely give a very
tiny contribution to $P_0(t)$ accompanying with a very
high-frequency oscillation, whereas the dominant part is given by
\begin{equation}
P_0(t) \approx \frac{1}{2}+\frac{1}{2}\cos (4\frac{J^2}{U}t).
\end{equation}
It is obvious that $P_0(t) \approx 1$ if $ J t \ll U/J$. Therefore
at the large $U$ limit, the atom pair is dynamically stable within
a very large time scale.

\section{Dynamics in the large U limit}

In the strongly repulsive interaction limit of $U/J \gg 1$, it is
convenient to represent the BHH as
\[
\widehat{H}=\widehat{H}_U+\widehat{H}_J,
\]
where $\widehat{H}_U$ and $\widehat{H}_J$ are the on-site
interaction part and tunnelling part respectively. We may solve
the following dynamical evolution equation
\begin{equation}
\left| \psi (t)\right\rangle =e^{-i\widehat{H}_Jt-i\widehat{H}_Ut}\left|
\psi _0\right\rangle
\end{equation}
approximately by treating $\widehat{H}_J$ as a perturbation. In order to do
that, we formally represent $\left| \psi (t)\right\rangle $ as
\begin{equation}
\left| \psi (t)\right\rangle =e^{-i\widehat{H}_Ut}U(t)\left| \psi
_0\right\rangle
\end{equation}
with $U(t)$ defined as $U(t)=e^{i\widehat{H}_Ut}e^{-i\widehat{H}_Jt-i%
\widehat{H}_Ut}$. It is easy to check that $U(t)$ fulfills the following
motion of equation
\[
\frac \partial {\partial t}U(t)=-i\widehat{H}_J(t)U(t)
\]
with
\[
\widehat{H}_J(t)=e^{i\widehat{H}_Ut}\widehat{H}_Je^{-i\widehat{H}_Ut}.
\]
Therefore the formal solution of $U(t)$ is given by
\begin{eqnarray*}
U(t) &=&T\exp \left[ -i\int_0^tdt_1\widehat{H}_J(t_1)\right] \\
&=&\sum_{n=0}^\infty \frac{\left( -i\right) ^n}{n!}\int_0^tdt_1\cdots
\int_0^tdt_nT\left[ \widehat{H}_J(t_1)\cdots \widehat{H}_J(t_n)\right] ,
\end{eqnarray*}
where $T$ is the time-ordering operator. By using the above formula, now we
can calculate $\left\langle \psi (0)\right. \left| \psi (t)\right\rangle $
perturbatively. Up to the second order, we get
\begin{eqnarray}
&&\left\langle \psi (0)\right. \left| \psi (t)\right\rangle  \nonumber \\
&=&\left\langle \psi _0\right| e^{-i\widehat{H}_Ut}\left| \psi
_0\right\rangle -  \nonumber \\
&&\left\langle \psi _0\right| e^{-i\widehat{H}_Ut}\int_0^tdt_1%
\int_0^{t_1}dt_2\widehat{H}_J(t_1)\widehat{H}_J(t_2)\left| \psi
_0\right\rangle . \label{PB}
\end{eqnarray}
From the definition of $\left| \psi (t)\right\rangle ,$ we see that it
should be normalized. However, in the scheme of perturbation calculation,
the normalized condition is not automatically fulfilled. In order to
overcome this problem and get a well-defined $P_0(t)$, we need use a
normalized $\left| \widetilde{\psi }(t)\right\rangle $ which is defined as
\begin{equation}
\left| \widetilde{\psi }(t)\right\rangle =\frac{\left| \psi (t)\right\rangle
}{\sqrt{\left\langle \psi (t)\right. \left| \psi (t)\right\rangle }}.
\end{equation}
Therefore in the scheme of perturbation expansion, the probability $P_0(t)$
is defined by
\begin{equation}
P_0(t)=\left| \left\langle \psi (0)\right. \left| \widetilde{\psi }%
(t)\right\rangle \right| ^2. \label{Pt}
\end{equation}
\begin{figure}[tbp]
\includegraphics[width=9cm]{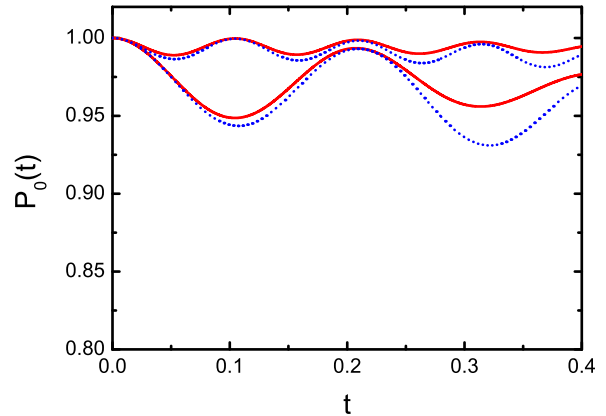}\newline
\caption{(Color online) Time evolution of $P_0(t)$ for the
six-site-six-boson case. The comparison between the solution given
by perturbation theory and the exact solution from numerical
computation is shown. The analytical results are denoted by the
dotted lines and the numerical result by solid lines. The top two
lines correspond to $U/J=60$ and the other two lines correspond to
$U/J=30$. The time is in units of $J$.}\label{Figure2}
\end{figure}

For a given initial state $\left| \psi _0\right\rangle $ represented in a
basis of Fock states
\begin{equation}
\left| \psi _0\right\rangle =\left| n_1,n_2,...,n_i,...,n_L\right\rangle
\end{equation}
with $n_i=0$ or $2,$ it is easy to get $\widehat{H}_U$ $\left|
\psi _0\right\rangle =\frac N2U$ $\left| \psi _0\right\rangle .$
For convenience, we first calculate the case with $\left| \psi
_0\right\rangle =\left| 2,0,2,0...,2,0\right\rangle $. By using
the formula (\ref{PB}), we get
\begin{eqnarray}
&&\left\langle \psi (0)\right. \left| \psi (t)\right\rangle  \nonumber \\
&=&e^{-i\frac N2Ut}\left[
1-i\frac{2NJ^2t}{U}+\frac{2NJ^2}{U^2}\left( e^{ iUt}-1\right)
\right]. \label{psi0t}
\end{eqnarray}
Therefore we have
\begin{eqnarray}
P_0(t) &=&A^{-1}\left[ 1-\left( \frac{4NJ^2}{U^2}-\frac{8N^2J^4}{U^4}\right)
\left( 1-\cos Ut \right) +\right.\nonumber \\
&&\left. \frac{4N^2J^4t^2}{U^2}-\frac{8N^2J^4t}{U^3 }\sin
Ut\right] \label{p0t2}
\end{eqnarray}
where $A=\left\langle \psi (t)\right. \left| \psi (t)\right\rangle
$ is the normalized constant.
We note that, in the scheme of the second-order perturbation
theory, formula (\ref{psi0t}) has the same form for different
initial configurations such as $\left| \psi _0\right\rangle
=\left| 2,0,2,0,0,2,0,0,0\right\rangle $ and $\left| \psi
_0\right\rangle =\left| 2,0,2,0,2,0,0,0,0\right\rangle $.
Actually, the formula (\ref{psi0t}) is valid for all the initial
configurations where the atoms pairs are not neighboring to each
others. For the other kind of initial configurations, it is
straightforward to calculate $P_0(t)$ by formula (\ref{Pt}) and
(\ref{PB}).

In order to check the validity of the perturbation theory, we plot
the $P_0(t)$ obtained by the perturbation theory for the two-site
problem in Fig. \ref {Figure1} and make a rough comparison between
the perturbation result and the exact result given in Eq. (\ref
{2exact}). Specially, for the simplest two-site case with $\left|
\psi _0\right\rangle =\left| 2,0 \right\rangle$ or
$\left|0,2\right\rangle$, we have
\begin{eqnarray}
P_0(t)&=&{A}^{-1}\left[1-\left(\frac{4J^2}{U^2}-\frac{8J^4}{U^4}\right)
\left(1-\cos Ut\right)
+\frac{4J^4 t^2}{ U^2} -\right. \nonumber \\
&&\left. \frac{8J^4t}{U^3}\sin Ut \right],
\end{eqnarray}
where the normalization constant $A$ is given by
\begin{eqnarray}
A=1+8J^4\left[\frac{2}{U^4}\left(1-\cos
Ut\right)-\frac{2t}{U^3}\sin Ut+\frac{t^2}{U^2}\right].
\end{eqnarray}
As is shown in Fig. \ref{Figure1}, the larger the $U$, the better
the exact and the approximate solution fit.

\begin{figure}[tbp]
\includegraphics[width=9cm]{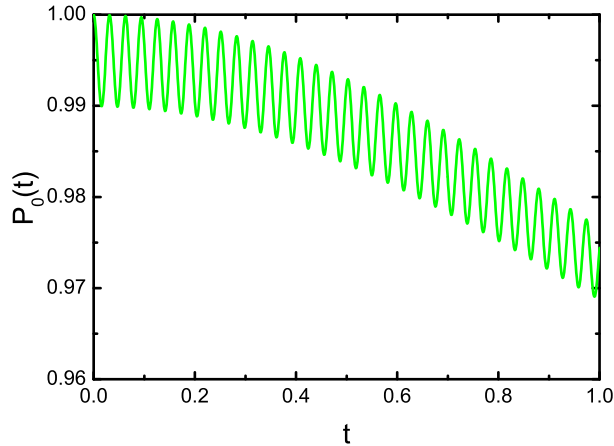}
\newline
\caption{Time evolution of $P_0(t)$ for the 50-particle system
with initial state $\left| 2,0,2,0,...,2,0,2,0\right\rangle $
versus $t$ in units of $J$ with the corresponding on-site
interaction $U/J=200$.} \label{Figure3}
\end{figure}

It is natural to use the formula (\ref{Pt}) to deal with the
many-particle system. From (\ref{psi0t}), we see that the
perturbation theory should work well as long as $ NJ^2t/U \ll 1$.
In order to compare with the exact result by numerical exact
diagonalization method in next section, we apply (\ref{p0t2}) to
calculate the time evolution of $P_0(t)$ for a finite-size system
with an initial state $\left|\psi_0\right\rangle=\left
|2,0,2,0,2,0 \right \rangle.$ Our result is shown in Fig. \ref
{Figure2}. We can see that at large $U$ limit, the perturbation
theory works quite well. Our results show that $\psi(t)$ decays
slower for a larger $U$. For a larger system with $N=50$ and the
initial state given by $\left| 2,0,2,0,...,2,0,2,0\right\rangle $,
the time evolution of $P_0(t)$ shows similar behaviors as that of
the 2- and 6-particle systems. As shown in Fig. \ref {Figure3},
$P_0(t)$ decays slowly accompanying with a high-frequency but very
narrow oscillation. These results show that, when $U$ is large
enough, $\psi(t)$ mainly stays in the initial state within a large
time scale. The larger the U is, the more obvious the case.

\section{Numerical result}

The exact diagonalization (ED) method is very powerful from the
viewpoint that it can give out all information of the system
exactly.
To solve the matrix eigenvalue problem of the Hamiltonian, it is
convenient to work in a basis of Fock states
\begin{equation}
\left| \alpha \right\rangle =\left| \{n_1,n_2,...,n_L\}_\alpha \right\rangle
,
\end{equation}
with $n_i=0,1,2,\cdots $ and the index $\alpha =1,...,D$ labels
the different compositions of occupation numbers for a fixed total
particle number $\sum_{i=1}^Ln_i=N$. Through the exact
diagonalization algorithm we compute exactly all the eigenvalues
$E_n$ and the corresponding eigenstates $\phi _n$ of the
Hamiltonian (\ref{BHH}). Then we can calculate the dynamic
evolution of $P_0(t)$ by using Eq. (\ref{p0t}).
To give a concrete example and compare with the result obtained by
the perturbation theory in Sec. III, we first consider an initial
state $\left|\psi_0\right\rangle=\left |2,0,2,0,2,0 \right
\rangle$. Our result is shown in Fig. \ref {Figure2}. We can see
that at large $U$ limit, the perturbation theory works quite well.

We note that what was measured in the experiment \cite{Winkler} is
the number of remaining pairs after a variable hold time. To
understand the experimental result qualitatively, we calculate the
time-dependent normalized number of the atom pairs which can be
represented as
\begin{equation}
\bar{N}_p(t)=\sum_{\alpha =1}^D\frac{N_p(\alpha) \left| c_\alpha
(t)\right| ^2} M,  \label{Npt}
\end{equation}
where $c_\alpha (t)$ is given by
\begin{equation}
c_\alpha (t)=\sum_{n=1}^De^{-iE_nt}\left\langle \alpha |\phi
_n\right\rangle \left\langle \phi_n|\psi_0\right \rangle ,
\end{equation}
$N_p(\alpha)$ denotes the number of atom pairs in $\left| \alpha
\right\rangle $ and $M=N/2$ is the total number of initial atom
pairs. Using the formula (\ref{Npt}), we can calculate the
normalized atom pair number for different on-site interaction $U$
and different initial state $\left| \psi _0\right\rangle $.
\begin{figure}[tbp]
\includegraphics[width=9cm]{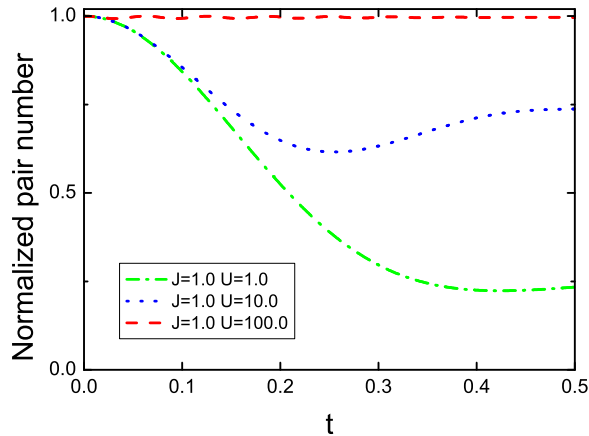}\newline
\caption{(Color online) Time evolution of normalized atom pair
number in a nine-site optical lattice for $U/J=1.0$, $10.0$ and
$100.0$. The time is in units of $J$ and the initial state is
chosen as $\left| 2,2,2,0,0,0,0,0,0\right\rangle $.
}\label{Figure4}
\end{figure}

For simplicity, let us consider a one-dimensional (1D) optical
lattice with $L=9$ and $M=3$, i.e. three boson atom pairs loaded
into an optical lattice of nine sites. The time dependent
normalized pair numbers for various interaction strengths are
plotted in Fig. \ref {Figure4} with the initial state given by $
\left| 2,2,2,0,0,0,0,0,0\right\rangle $. Evidently, the atom pairs
in the case of $U=100.0$ stay longer than that in the case of
$U=1.0$. It is clear that the large on-site interaction $U$
effectively suppresses the dynamic dissociation of the initial
pairs, which is in qualitative agreement with the experiment
reported in Ref. \cite{Winkler}.

\begin{figure}[tbp]
\includegraphics[width=9cm]{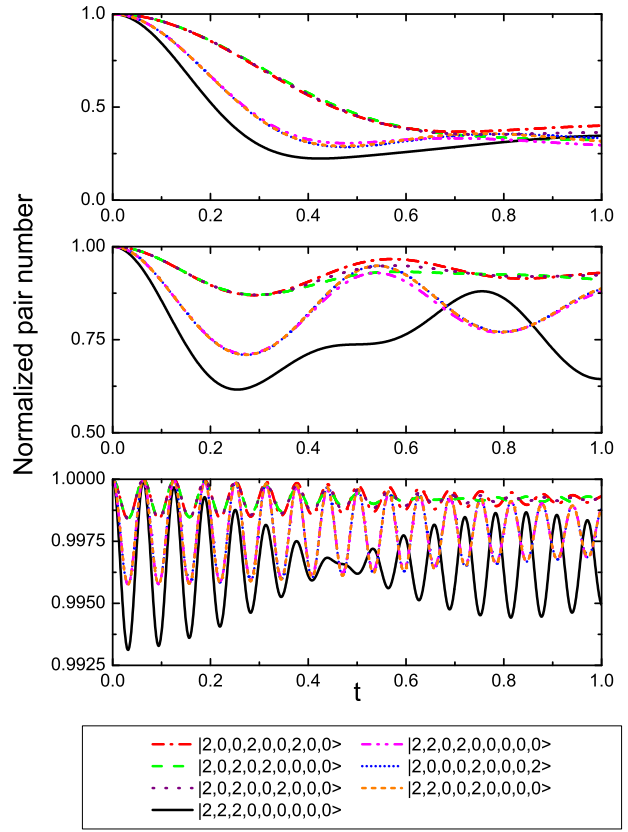}\newline
\caption{(Color online) Time evolution of the normalized atom pair
number for different on-site interaction $U$ and different initial
states $\left| \psi _0\right\rangle $. From top to bottom the
on-site interaction $U/J$= 1.0, 10.0, and 100.0, respectively. In
each graph, there are seven lines corresponding to seven different
initial states. The initial states are denoted by different lines
shown in the table at the bottom. The time is in units of $J$.
}\label{Figure5}
\end{figure}

Next we study how the dynamics of the repulsively bound pairs is
affected by the choice of the initial state. For the case with
three bound pairs loaded into an optical lattice with nine sites,
there are $C_9^3=84$ different configurations. Taken the
translational symmetry of the system into account, there are in
fact only seven different configurations. For different initial
state $\left| \psi _0\right\rangle $, we calculate the normalized
atom pair number and plot them in Fig. \ref{Figure5}. According to
Fig. \ref{Figure5}, although for different initial state $\left|
\psi _0\right\rangle $ the corresponding dynamics of the system is
similar to each other, there are still some minor differences
depending on whether the initial pairs are close together or not.
For initial states $\left| 2,0,2,0,2,0,0,0,0\right\rangle $,
$\left| 2,0,2,0,0,2,0,0,0\right\rangle $, and $\left|
2,0,0,2,0,0,2,0,0\right\rangle $ in which each pair is apart from
the others, the time evolutions are almost the same. This is
consistent with our result by perturbation theory. According to
the second-order perturbation theory, the time evolution of
$P_0(t)$ for these initial states are same. Similarly, the time
evolution for initial states $\left|
2,2,0,2,0,0,0,0,0\right\rangle $, $\left|
2,2,0,0,2,0,0,0,0\right\rangle $, and $\left|
2,2,0,0,0,2,0,0,0\right\rangle $ are almost the same for different
on-site interaction $U$ as is shown in Fig. \ref{Figure5}. We
notice that the initial states with bound pairs repelling each
other are dynamically stabler than the states with all atom pairs
initially occupying neighboring sites. We also investigate the
case in which there are three boson atom pairs loaded into a
two-dimensional $3\times 3$ optical lattice. The results are
similar to the one-dimensional case as shown in Fig. 4 and Fig. 5.

\begin{figure}[tbp]
\includegraphics[width=8cm]{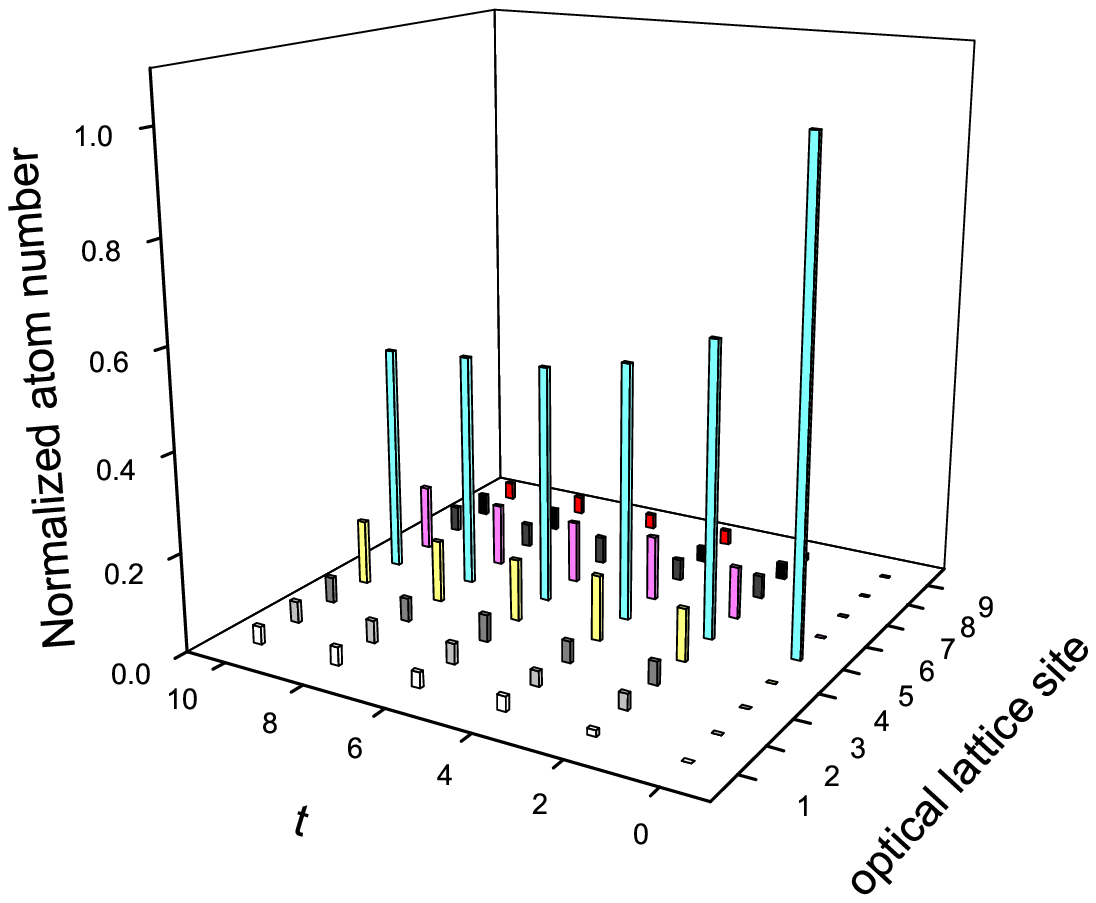}
\includegraphics[width=8cm]{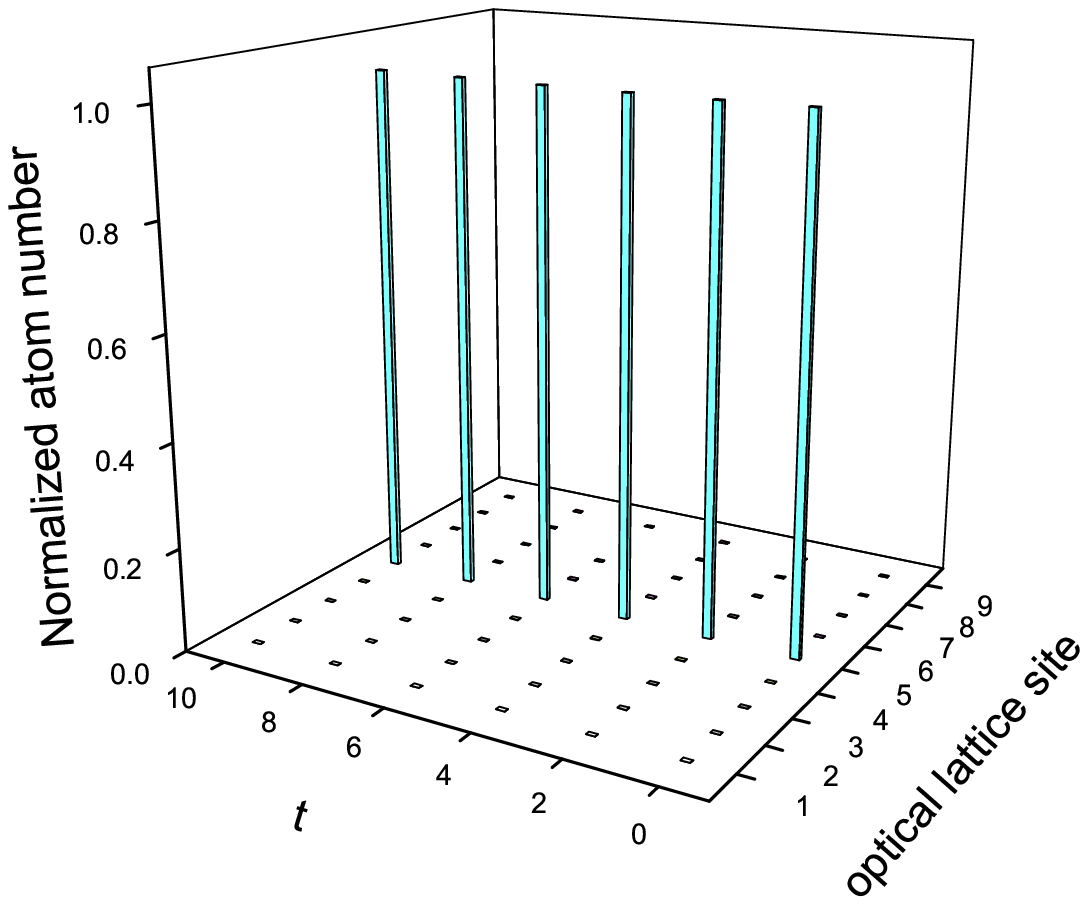}
\newline
\caption{Time evolution of the normalized atom number on each site
of the optical lattice. We choose the initial state $\left| \psi_0
\right\rangle $ as $\left| 0,0,0,0,6,0,0,0,0\right\rangle $. The
graph on the top corresponds to $U/J$=1.0 and the graph on the
bottom to $U/J$=10.0. The time is in units of $J$. }
\label{Figure6}
\end{figure}

From above discussions, we could understand the stable atom-pair
state due to the strongly repulsive interaction as a dynamically
stable state. The physical origin is essentially that the energy
conservation does not allow the particle to tunnel into the
neighboring sites to save the repulsive interaction unless the
obtained tunneling energy is comparable to the loss of repulsive
energy \cite{BlochRMP}. From theoretical point of view, a similar
dynamically stable state is expected to appear for the case in
which the initial state composed of three or more atoms. In order
to clarify this point, we investigate an example case in which six
atoms were initially loaded into only one site of a 1D optical
lattice with nine sites. The initial state is chosen as $\left|
0,0,0,0,6,0,0,0,0\right\rangle $. Given the initial state, we then
determine the dynamic evolution of the initial state on the whole
lattice, \textit{i.e.}, the spatial distribution of the normalized
atom number on each site at different time point for different
on-site interaction. The time-dependent normalized atom number on
the $j$-th site can be calculated by
\begin{equation}
\bar{n}_j(t)=\sum_{\alpha=1}^D\frac{\left| c_\alpha(t)\right|
^2n_j^{\alpha}} N,
\end{equation}
where $n_j^{\alpha}=\left\langle \alpha \right| \hat{n}_j \left|
\alpha \right\rangle$. The spatial distributions provide us useful
information for the expansion dynamics of a condensate loaded in
the optical lattice. At an incremental sequence of time points, we
calculate the normalized atom number at each site. The results are
shown in Fig. \ref{Figure6}. For a small $U$ $(U=J)$, the atoms
initially trapped in a site tend to tunnel to the neighboring
sites and thus lead to the decrease of the normalized atom number
in the initial site. However, when the on-site $U$ is large
enough, the initially trapped atoms are dynamically frozen and the
expansion to the neighboring sites is almost completely
suppressed.

\section{summary}
In this paper, we theoretically investigate the dynamical
stability of atom pairs induced by the strongly repulsive
interaction in an optical lattice both analytically and
numerically. Firstly, in the large $U$ limit, we calculate the
time evolution of an initial system composed of atom pairs in a
scheme of the perturbation theory. The analytical results show
that the initial state of atom pairs is dynamically stabler for
larger repulsive interaction. Then by the exact diagonaliztion
method, we numerically study the stability of atom pairs in a
finite-size system with few particles by calculating the time
evolution of the normalized atom pair number. Our results show
that the initial state of atom pairs are dynamically stable and
the dissociation of atom pairs is greatly suppressed when the
strength of the on-site interaction is large. We also compare our
numerical result with the result based on the perturbation theory
and find out that they fit quite well in the large $U$ limit. Our
results imply that the experimentally observed repulsively bound
pairs can be understood as a dynamically stable state.

\begin{acknowledgments}
This work is supported by NSF of China under Grant No. 10574150
and programs of Chinese Academy of Sciences. The authors wish to
thank Ninghua Tong for useful discussions.
\end{acknowledgments}

\end{document}